\documentclass{ws-ijgmmp}

\usepackage{url}

\begin{document}

\markboth{M. Planat, A. Giorgetti, F. Holweck and M. Saniga}
{Quantum contextual finite geometries from \textit{dessins d'enfants}}

%
\catchline{}{}{}{}{}
%

\title{Quantum contextual finite geometries from \textit{dessins d'enfants}}

\author{MICHEL PLANAT}

\address{Institut FEMTO-ST/MN2S, CNRS, 15B Avenue des Montboucons - 25030 Besan\c con, France \\
\email{ michel.planat@femto-st.fr} }

\author{ALAIN GIORGETTI}
\address{Institut FEMTO-ST/DISC, Universit\'e de Franche-Comt\'e, 16 route de Gray,\\ F-25030 Besan\c con, France\\
\email{ alain.giorgetti@femto-st.fr} }

\author{FR\'ED\'ERIC HOLWECK}
\address{Laboratoire IRTES/M3M, Universit\'e de Technologie de Belfort-Montb\'eliard, \\F-90010
Belfort, France \\
\email{ frederic.holweck@utbm.fr} }

\author{METOD SANIGA}
\address{Institute of Discrete Mathematics and Geometry, Vienna University of Technology, Wiedner Hauptstrasse 8-10, A-1040 Vienna, AUSTRIA \footnote{Astronomical Institute, Slovak Academy of Sciences, SK-05960 Tatransk\'a Lomnica,
Slovak Republic} \\
\email{ msaniga@astro.sk} }

\maketitle

\begin{history}
\received{(Day Month Year)}
\revised{(Day Month Year)}
\end{history}

\begin{abstract}
We point out an explicit connection between graphs drawn on compact Riemann surfaces defined over the field $\bar{\mathbb{Q}}$ of algebraic numbers --- so-called Grothendieck's {\it dessins d'enfants} --- and a wealth of distinguished point-line configurations. These include simplices, cross-polytopes, several notable projective configurations,  a number of multipartite graphs and some `exotic' geometries.  Among them, remarkably, we find not only those underlying Mermin's magic square and magic pentagram, but also those related to the geometry of two- and three-qubit Pauli groups. Of particular interest is the occurrence of all the three types of slim generalized quadrangles, namely GQ(2,1), GQ(2,2) and GQ(2,4), and a couple of closely related graphs, namely the Schl\"{a}fli and Clebsch ones. These findings seem to indicate that {\it dessins d'enfants} may provide us with a new powerful tool for gaining deeper insight into the nature of finite-dimensional Hilbert spaces and their associated groups, with a special emphasis on contextuality.
\end{abstract}



\section{Introduction}

If one draws a (connected) graph --- a particular set of vertices and edges --- on a smooth surface, then such graph inherits extra local/combinatorial and global/topological features from the surface. If the latter is, for example, a (compact) complex one-dimensional surface --- a Riemann surface, then the combinatorics of edges is encapsulated by a two-generator permutation group and the Riemann surface happens to be definable over the field $\bar{\mathbb{Q}}$ of algebraic numbers. This observation is central to the concept of {\it dessins d'enfants} (or child's drawings) as advocated by Grothendieck in his {\it Esquisse d'un programme} (made available in 1984 following his {\it Long March} written in 1981) in the following words:
{\it In the form in which Belyi states it, his result essentially says that every algebraic curve defined over a number field can be obtained as a covering of the projective line ramified only over the points 0, 1 and $\infty$. The result seems to have remained more or less unobserved. Yet it appears to me to have considerable importance. To me, its essential message is that there is a profound identity between the combinatorics of finite maps on the one hand, and the geometry of algebraic curves defined over number fields on the other. This deep result, together with the algebraic interpretation of maps, opens the door into a new, unexplored world - within reach of all, who pass by without seeing it} \cite[Vol. 1]{Grothen}, \cite{Schneps1}.

Our aim is to show that Grothendieck's {\it dessins d'enfants} (see, e.\,g., \cite{zapp2003, Lando2004} as well as \cite{Koch2010}) have, as envisaged in \cite{PlanatFQXi}, great potential to become a proper language for a deeper understanding of various types of sets of Hermitian operators/observables that appear in finite-dimensional quantum mechanical settings and for furnishing a natural explanation why eigenvalues of these operators are regarded as the only available tracks in associated measurements. The main justification of our aim is provided by the fact that {\it dessins} lead very naturally to {\it already}-discovered finite geometries underlying quantum contextuality (like the grid, GQ(2,\,1), behind Mermin's magic square and/or an ovoid of PG(3,\,2) behind Mermin's magic pentagram) and also to those underlying commutation relations between elements of the two-qubit Pauli group (the generalized quadrangle of order two, GQ(2,\,2), its geometric hyperplanes and their complements, see, e.\,g., \cite{Planat2007}).



The paper is organized as follows. Section \ref{dessins} gathers some basic
knowledge about {\it dessins d'enfants}, their permutation group and topology,
their isomorphism with conjugacy classes of subgroups of finite index of the
cartographic group $C_2^+$, as well as about associated Belyi functions.
Section \ref{square} focuses on a rather elementary application of our ideas by
interpreting Bell's theorem about non-locality in terms of the geometry  as
simple as a square/quadrangle, which is found to be generated by four distinct
{\it dessins} defined over the field $\mathbb{Q}[\sqrt{2}]$.
Section \ref{catalog}, the core one, starts with a complete catalog of all
connected geometries induced by {\it dessins} having up to $12$ edges and a
sketch of important ones with more edges. In the subsequent subsections, we
analyze in detail the non-trivial cases by selecting, whenever possible, a {\it
dessin} of genus zero and having the smallest number of faces. As in most cases
the edges of {\it dessins} dealt with admit labeling by two- or three-qubit
observables, on our way we not only encounter already recognized
quantum-relevant finite geometries like the Fano plane, the grid GQ(2,\,1), the
Petersen graph, the Desargues configuration and the generalized quadrangle
GQ(2,\,2), but find a bunch of novel ones, some already surmised from different
contexts, starting from the Pappus $9_3$-configuration and the Hesse $(9_4,
12_3)$-configuration ({\it aka} the affine plane AG(2,\,3)) to arrive at the
generalized quadrangle GQ(2,\,4) --- and its close siblings, the Clebsch and
Schl\"{a}fli graphs -- known to play a role in the context of the
black-hole--qubit correspondence \cite{Levay2009}.
Section 5 is reserved for concluding remarks.



\section{\textit{Dessins d'enfants} and the Belyi theorem}
\label{dessins}


\subsection*{Dessins d'enfants and their symmetry groups}

A {\it map} is a graph drawn on a surface --- a smooth compact orientable
variety of dimension two --- such that its vertices are points, its edges are
non-intersecting arcs connecting the vertices, and the connected
components of its complement, called faces, are
homeomorphic to open disks of $\mathbb{R}^2$. They may exist multiple edges as well as loops, but the graph has to be connected. Denoting the number of vertices, edges and faces by $V$, $E$ and $F$, respectively, the genus $g$ of the map follows from Euler's formula $V-E+F=2-2g$.
A map can be generalized to a bicolored map. The latter is a map whose vertices are colored in black and white in such a way that the adjacent vertices have always the opposite color; the corresponding segments are the edges of the bicolored map. The Euler characteristic now reads $2-2g=B+W+F-n$, where $B$, $W$ and $n$ stands for the number of black vertices, the number of white vertices and the number of edges, respectively. Given a bicolored map with $n$ edges labeled from 1 to $n$, one can associate with it a permutation group $P=\left\langle \alpha, \beta \right\rangle$ on the set of labels such that a cycle of $\alpha$ (resp. $\beta$) contains the labels of the edges incident to a black vertex (resp. white vertex),
taken, say, in the clockwise direction around this
vertex; thus, there are as many cycles in $\alpha$ (resp. $\beta$)  as there are black (resp. white vertices),
and the degree of a vertex is equal to the length of the corresponding
cycle. An analogous cycle structure for the faces follows from the permutation $\gamma$ satisfying $\alpha\beta\gamma=1$.

Bicolored maps (allowed to have any valency for their vertices) are in
one-to-one correspondence with hypermaps \cite{Walsh}. They correspond to the
conjugacy classes of subgroups of finite index of the free group on two
generators $H_2^+=\left\langle  \rho_0,\rho_1 \right\rangle$.
The number of hypermaps with $n$ half-edges is given by the sequence
{\it A057005} in the OEIS~\cite{A057005}. For $n= 1,\ldots,7$ these
numbers are $1$, $3$, $7$, $26$, $97$, $624$ and $4163$.
Hypermaps are, of course, allowed to have any valency for their
vertices.

We consider bicolored maps where the valency of white vertices is $\leq 2$.
They correspond to hypermaps of the so-called pre-clean type.
As already observed  by Grothendieck himself, who called them {\it dessins
d'enfants} (child's drawings)~\cite{Schneps1}--\cite{Girondo}, these bicolored
maps on connected oriented surfaces are unique in the sense that they are in
one-to-one correspondence with conjugacy classes of subgroups of finite index of
the triangle group, also called {\it cartographic} group
\begin{equation}
C_2^+=\left\langle  \rho_0,\rho_1|\rho_1^2=1 \right\rangle.
\label{cartographic}
\end{equation}
The existence of associated {\it dessins} of prescribed properties can thus be
straightforwardly checked from a systematic enumeration of conjugacy classes of
$C_2^+$; with the increasing $n>0$ the number of such {\it dessins} grows quite
rapidly \[1, 3, 3, 10, 15,   56, 131, 482, 1551, 5916,   22171, 90033,
370199,\cdots\]

A {\it dessin} $\mathcal{D}$ can be ascribed a signature $s=(B,W,F,g)$ and the full information about it can be recovered from the structure of the generators of its permutation group $P$ (also named the passport in \cite{Lando2004,Zvonkin}) in the form $[C_{\alpha},C_{\beta},C_{\gamma}]$, where the entry $C_i$, $i \in \{\alpha,\beta,\gamma\}$ has factors $l_i^{n_i}$, with $l_i$ denoting the length of the cycle and $n_i$ the number of cycles of length $l_i$.

\subsection*{Belyi's theorem}

Given  $f(x)$, a rational function of the complex variable $x$, a {\it critical point} of $f$ is a root of its derivative and a {\it critical value} of $f$ is the value of $f$ at the critical point. 
Let us define a so-called {\it Belyi function} corresponding to a
dessin $\mathcal{D}$ as a rational function $f(x)$ of degree $n$ if $\mathcal{D}$ may be embedded into
the Riemann sphere $\hat{\mathbb{C}}$ in such a way that (i) the black vertices are the roots of the equation $f(x)=0$ with the multiplicity of each root being equal to the degree of the corresponding (black) vertex, (ii) the white vertices are the roots of the equation $f(x)=1$ with the multiplicity of each root being equal to the degree of the corresponding (white) vertex,
(iii) the bicolored map is the preimage of the segment $[0,1]$, that is $\mathcal{D}=f^{-1}([0,1])$,
(iv) there exists a single pole of $f(x)$, i.\,e. a root of the equation $f(x)=\infty$, at each face, the multiplicity of the pole being equal to the degree of the face, and, finally, 
(v) besides $0$, $1$ and $\infty$, there are no other critical values of $f$ \cite{Zvonkin}.

It can be shown that to every $\mathcal{D}$ there corresponds a Belyi function $f(x)$ and that this function is, up to a linear fractional transformation of the variable $x$, unique. It is, however, a highly non-trivial task to find/calculate the Belyi function for a general {\it dessin}.

\subsection*{Finite geometries from dessins d'enfants}

An issue of central importance for us is the fact that one can associate with a dessin $\mathcal{D}$ a {\it point-line incidence geometry}, $G_{\mathcal{D}}$, in the following way. A point of $G_{\mathcal{D}}$ corresponds to an edge of $\mathcal{D}$. Given a dessin $\mathcal{D}$, we want its permutation group $P$ to preserve the collineation of the geometry $G_{\mathcal{D}}$. We first ask that every pair of points on a line shares the same stabilizer in $P$. Then, given a subgroup $S$ of $P$ which stabilizes a pair of points, we define the point-line relation on $G_{\mathcal{D}}$ such that two points on the same line share the same stabilizer. The lines on a geometry are distinguished by their (isomorphic) stabilizers acting on different $G$-sets. This construction allows to assign finite geometries $G_{\mathcal{D}^i}$ to a dessin $\mathcal{D}$, $i=1,\cdots,m$ with $m$ being the number of non-isomorphic subgroups $S$ of $P$ that stabilize a pair of elements \footnote{Our definition follows an example of the action on the Fano plane of a permutation group of order $|PSL(2,7)|=168$ associated with a tree-like {\it dessin} (of the relevant cycle structure) given in  \cite[(a), vol. 2,  p. 17 and p. 50]{Schneps1}.}.  As a slight digression we mention that, presumably, this action of the group $P$ of a dessin $\mathcal{D}$ on the associated geometry $G_{\mathcal{D}}$ is intricately linked with the properties of the absolute Galois group $\Gamma=\mbox{Gal}(\bar{\mathbb{Q}}/\mathbb{Q})$, which is the group of automorphisms of the field $\bar{\mathbb{Q}}$ of algebraic numbers.  Although $\Gamma$ is known to act faithfully on $\mathcal{D}$ \cite{Schneps1,Girondo},
its action on $G_{\mathcal{D}}$  must be rather non-trivial because the map from $\mathcal{D}$ to $G_{\mathcal{D}}$ is non injective.  Further work is necessary along this line of thoughts to clarify the issue.

Using a computer program, we have been able to completely catalogize incidence geometries associated with {\it dessins} featuring up to 12 edges, and also found several {\it dessins} of higher rank that produce distinguished geometries. The results of our calculations are succinctly summarized in Tables 1 and 2.
The subsequent sections provide a detailed account of a variety of {\it dessins} computed, their corresponding  point-line incidence geometries, and, what is perhaps most important, how these relate to the physics of quantum observables of multiple-qubit Pauli groups and related quantum paradoxes. In other words, we shall give a more exhaustive and rigorous elaboration of the ideas first outlined in a short essay-like treatise \cite{PlanatFQXi}.

\bigskip 

%
\begin{table}[ht]
\begin{center}
\begin{tabular}{|r|r|r|r|r|r|}
\hline
index & name & vertices & edges & triangles & squares \\
\hline
$3$ & $2$-simplex (triangle) & $3$ & $3$ & $1$ & $0$ \\
\hline
$4$ & $3$-simplex (tethahedron) & $4$ & $6$ & $4$ & $0$ \\
$$ & square/quadrangle & $4$ & $4$ & $0$ & $1$ \\
 \hline
 $5$ & $4$-simplex ($5$-cell) & $5$ & $10$ & $10$ & $0$ \\
 \hline
 $6$ & $5$-simplex  & $6$ & $15$ & $20$ & $0$ \\
 $$ & $3$-orthoplex (octahedron)  & $6$ & $12$ & $8$ & $3$ \\
 $$ & bipartite graph $K(3,3)$   & $6$ & $9$ & $0$ & $9$ \\
 \hline
 $7$ & $6$-simplex  & $7$ & $21$ & $35$ & $0$ \\
 $$ & Fano plane ($7_3$)  & $7$ & $21$ & $7$ & $0$ \\
 \hline
 $8$ & $7$-simplex  & $8$ & $28$ & $56$ & $0$ \\
 $$ & $4$-orthoplex ($16$-cell)  & $8$ & $24$ & $32$ & $6$ \\
 $$ & completed cube $K(4,4)$  & $8$ & $16$ & $0$ & $36$ \\
 $$ & stellated octahedron   & $8$ & $12$ & $8$ & $0$ \\
  \hline
 $9$ & $8$-simplex  & $9$ & $36$ & $84$ & $0$ \\
 $$ & Hesse ($9_412_3$)  & $9$ & $36$ & $12$ & $0$ \\
 $$ & $K(3)^3$ & $9$ & $27$ & $27$ & $27$ \\
 $$ & Pappus ($9_3$)  & $9$ & $27$ & $9$ & $27$ \\
 $$ & ($3 \times 3$)-grid  & $9$ & $18$ & $6$ & $9$ \\
 \hline
 $10$ & $9$-simplex  & $10$ & $45$ & $120$ & $0$ \\
 $$ & $5$-orthoplex  & $10$ & $40$ & $80$ & $10$ \\
 $$  & bipartite graph $K(5,5)$  & $10$ & $25$ & $0$ & $100$ \\
 $$ & Mermin's pentagram  & $10$ & $30$ & $30$ & $15$ \\
 $$ & Petersen graph  & $10$ & $15$ & $0$ & $0$ \\
 $$ & Desargues ($10_3$)  & $10$ & $30$ & $10$ & $15$ \\
 \hline
 $11$ & $10$-simplex  & $11$ & $55$ & $165$ & $0$ \\
 \hline
 \hline
 $12$ & $11$-simplex & $12$ & $66$ & $220$ & $0$ \\
 $$ & $6$-orthoplex & $12$ & $60$ & $160$ & $15$ \\
 $$ & bipartite graph $K(6,6)$ & $12$ & $36$ & $0$ & $255$ \\
 $$ & threepartite graph $K(4,4,4)$ & $12$ & $48$ & $64$ & $108$ \\
 $$ & fourpartite graph $K(3,3,3,3)$ & $12$ & $54$ & $0$ & $54$ \\
 \hline
\end{tabular}
\label{geometries}
\normalsize
\caption{A catalog of connected point-line incidence geometries induced by {\it dessins d'enfants} of small index $\le 12$. For each geometry, when represented by its collinearity graph, we list the number of points, edges (line-segments joining two points), triangles and squares it contains. Here, $A$-simplices should be regarded as trivial because their {\it dessins} are star-like and associated Belyi functions are of a simple form $f(x)=x^A$, where $A$ is the multiplicity of the singular point at $x=0$.}

\end{center}
\end{table}
\normalsize

\begin{table}[ht]
\begin{center}
\begin{tabular}{|r|r|r|r|r|r|}
\hline
index & name & vertices & edges & triangles & squares \\
\hline
$15$ & Cremona-Richmond ($15_3$) (alias GQ(2,\,2))  & $15$ & $45$ & $15$ & $90$ \\
\hline
$16$ & Clebsch graph CG: $sp(10^1,2^5,-2^{10})$ & $16$ & $80$ & $0$ & $60$ \\
$$ & Shrikhande graph SG: $sp(6^1,2^6,-2^9)$ & $16$ & $48$ & $32$ & $12$ \\
\hline
$18$ &  $sp(8^1,0^9,-4^4,2^4)$& $18$ & $72$ & $48$ & $306$ \\
\hline
$20$ & $sp(0^{10},-8^1,8^1,-2^4,2^4)$& $20$ & $80$ & $0$ & $740$ \\
\hline
 $21$ & Kneser graph KG$_{(7,2)}$: $sp(10^1,3^6,-2^{14})$  & $21$ & $105$ & $35$ & $630$ \\
 $$ & $\mathcal{L}(IG(7,3,1)):~~~~ sp(4^1,-2^8,(1 \pm \sqrt{2})^6)$  & $21$ & $42$ & $14$ & $0$ \\
 \hline
 $22$ & $IG(11,5,2)$:~~$sp(\pm 5^1,\pm \sqrt{3}^{10})$  & $22$ & $55$ & $0$ & $55$ \\
 \hline
  $27$ & GQ(2,\,4): $sp(10^1,1^{20},-5^6)$  & $27$ & $135$ & $45$ & $1080$ \\
  $$ & Schl\"{a}fli graph SHG: $sp(16^1,4^6,-2^{20})$ & $27$ & $216$ & $720$ & $270$ \\
 $$ &  $sp(16^1,1^{16},-2^8,-8^2)$ & $27$ & $216$ & $504$ & $3024$ \\
 $$ &  $sp(16^1,4^2,1^{12},-2^8,-5^4)$ & $27$ & $216$ & $612$ & $1674$ \\
 \hline
 \end{tabular}
\label{geometries2}
\normalsize
\caption{A few (non-trivial) connected point-line incidence geometries induced by {\it dessins d'enfants} of  index greater than $12$. The spectra of (collinearity) graphs, denoted as $sp(\cdots, a^b,\cdots)$ where an eigenvalue $a$ is of multiplicity $b$, are also displayed. $IG(\nu,k,\lambda)$ means the incidence graph of a symmetric $2-(\nu,k,\lambda)$ design and $\mathcal{L}(\cdots)$ stands for the line graph.}
\end{center}
\end{table}
\normalsize

\section{The square geometry of Bell's theorem and the corresponding dessins}
\label{square}

{\it In a theory in which parameters are added to quantum mechanics to determine the results of individual measurements, without changing the statistical predictions, there must be a mechanism whereby the setting of one measuring device can influence the reading of another instrument, however remote \cite{Bell1964}}.

\subsection*{The square geometry of Bell's theorem}

\begin{figure}
\centering 
\includegraphics[width=7cm]{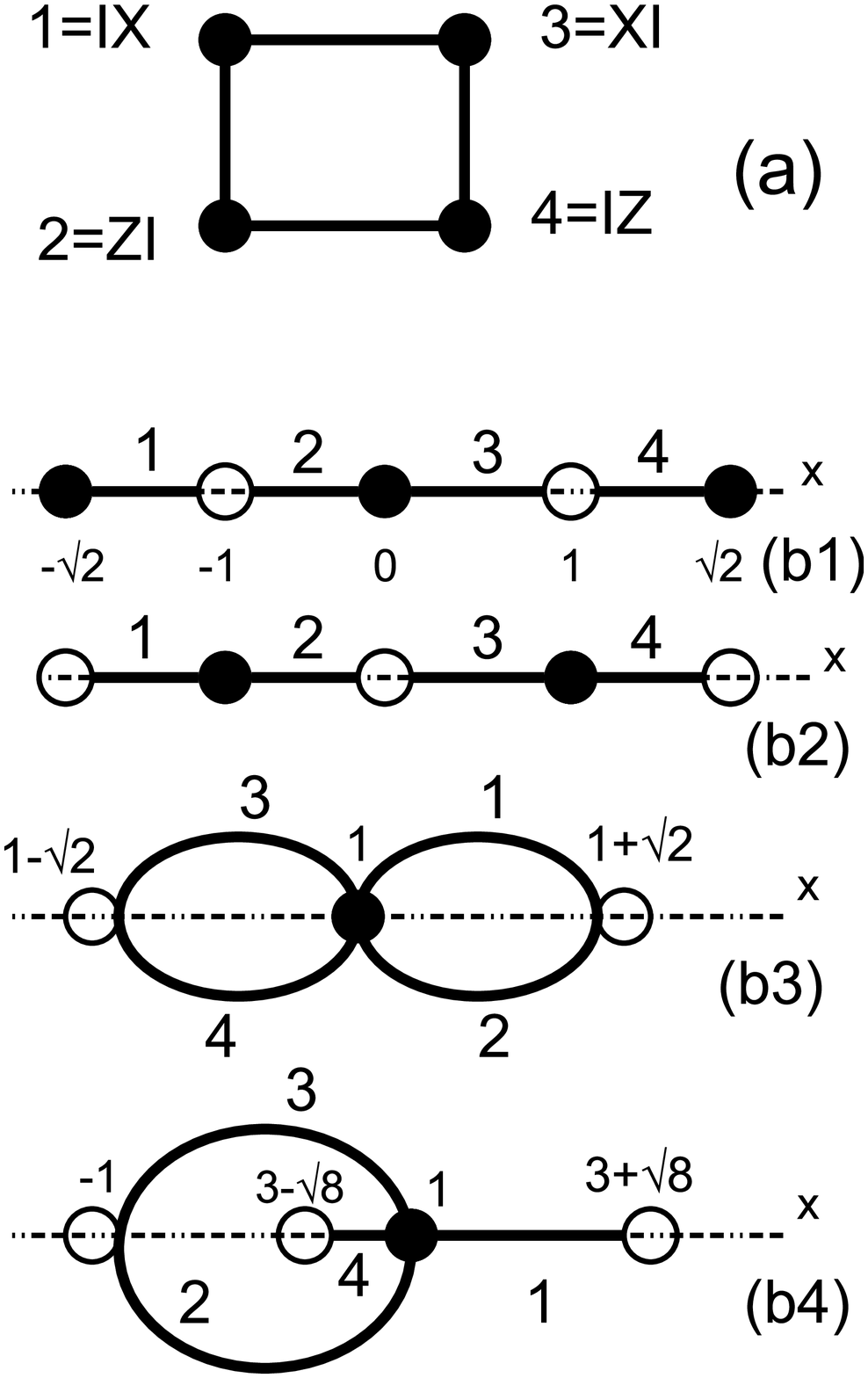}
\caption{A simple observable proof of Bell's theorem is embodied in the geometry of a (properly labeled) square (a) and four associated {\it dessins d'enfants}, ($b_1$) to ($b_4$). 
For each {\it dessin} an explicit labeling of its edges in terms of the four two-qubit observables is given. The (real-valued) coordinates of black and white vertices stem from the corresponding Belyi functions as explained in the main text.}
\label{fig1}
\end{figure}

Suppose we have four observables $\sigma_i$, $i = 1,2,3,4$, taking values in $\{-1,1\}$, of which Bob can measure $(\sigma_1,\sigma_3)$ and Alice $(\sigma_2,\sigma_4)$. The Bell-CHSH approach to quantum contextuality/non-locality consists of defining the number 
$$C=\sigma_2(\sigma_1+\sigma_3)+\sigma_4(\sigma_3-\sigma_1)=\pm 2$$
and observing the (so-called Bell-CHSH) inequality \cite[p. 164]{Peres}
$$|\left\langle \sigma_1\sigma_2 \right\rangle+\left\langle \sigma_2\sigma_3 \right\rangle+\left\langle \sigma_3\sigma_4 \right\rangle-\left\langle \sigma_4\sigma_1 \right\rangle|\le 2,$$
where $\left\langle  \right\rangle$ here means that we are taking averages over many experiments. This inequality holds for any dichotomic random variables $\sigma_i$ that are governed by a joint probability distribution. Bell's theorem states that the aforementioned inequality is violated if one considers quantum observables with dichotomic eigenvalues. An illustrative example is the following set of two-qubit observables
\begin{equation}
\sigma_1=IX,~\sigma_2=XI,~\sigma_3=IZ,~\sigma_4=ZI.
\label{quadruple}
\end{equation}
Here and below we use the notation $X$, $Y$ and $Z$ for the ordinary Pauli spin matrices and, e.\,g., $IX$ is a short-hand for $I \otimes X$ (used also in the sequel).
We find that
$$C^2=4*I +[\sigma_1,\sigma_3][\sigma_2,\sigma_4]=4 \left(\begin{array}{cccc} 1 &. & . &1 \\ . &1 & \bar{1} &.  \\ . &\bar{1} & 1&. \\ 1 &. & . &1   \end{array}\right)$$ 
has eigenvalues $0$ and $8$, both with multiplicity $2$  ($\bar{1} \equiv -1$). Taking the norm of the bounded linear operator $A$ as $||A||=sup (||A \psi||/||\psi||),~\psi \in \mathcal{H}$ ($\mathcal{H}$ being the corresponding Hilbert space), one arrives at the {\it maximal} violation of the Bell-CHSH inequality \cite[p. 174]{Peres}, namely
$||C||=2\sqrt{2}.$

The point-line incidence geometry associated with our four observables is one of the simplest, that of a square -- Fig.\ref{fig1}a; each observable is represented by a point and two points are joined by a segment if the corresponding observables commute. It is  worth mentioning here that there are altogether 90 distinct squares among two-qubit observables and as many as
 $30240$ when three-qubit labeling is employed  \cite{Planat2013}, each yielding a maximal violation of the Bell-CHSH inequality. 

\subsection*{Dessins d'enfants for the square  and their Belyi functions}

As it is depicted in Fig. \ref{fig1}, the geometry of square can be generated by four different {\it dessins}, $b_1,\cdots, b_4$, associated with permutations groups $P$ isomorphic to the dihedral group $D_4$ of order $8$.

The first {\it dessin} ($b_1$) has the signature $s=(B,W,F,g)=(3,2,1,0)$ and the permutation group $P=\left\langle (2,3),(1,2)(3,4)\right\rangle$ whose cycle structure reads $[2^1 1^2,2^2,4^1]$, i.e. one black vertex is of degree two, two black vertices have degree one, the two white vertices have degree two and the face has degree four. The corresponding Belyi function reads $f(x)=x^2(2-x^2)$. Its critical points are  $x \in \{-1, 1, 0\}$ and the corresponding critical values are $\{1,1,0\}$. The preimage of the value $0$ (the solutions of the equation $f(x)=0$) corresponds to the black vertices of the {\it dessin} positioned at $x \in \{-\sqrt{2},\sqrt{2},0\}$ and the preimage of the value $1$ (the solutions of the equation $f(x)=1$) corresponds to the white vertices at $x=\pm 1$.
The second {\it dessin} $(b_2)$ has $s=(2,3,1,0)$, $P=\left\langle (1,2)(3,4),(2,3)\right\rangle$ with $[2^2,2^1 1^2,4^1]$, and the Belyi function of the form $f(x)=(x^2-1)^2$.
The third {\it dessin} $(b_3)$ is characterized by $s=(1,2,3,0)$ and its $P=\left\langle (1,2,4,3),(1,2)(3,4)\right\rangle$ has the cycle structure $[4^1,2^2,2^11^2]$. The Belyi function may be written as
$f(x)=\frac{(x-1)^4}{4x(x-2)}$. As $f'(x)=\frac{(x-1)^3(x^2-2x-1)}{2(x-2)^2x^2}$, its critical points lie at $x=1$ (where $f(1)=0$)  and at $x=1\pm \sqrt{2}$ (where $f(1\pm \sqrt{2})=1$). 
Finally, the fourth {\it dessin} $(b_4)$ has $P=\left\langle (1,2,4,3),(2,3)\right\rangle$, the signature $s=(1,3,2,0)$ and cycle structure $[4^1,2^11^2,2^2]$. The Belyi function reads $f(x)=\frac{(x-1)^4}{16 x^2}$; hence, $f'(x)=\frac{(x-1)^3(x+1)}{8x^3}$. The critical points are at $x=-1$ (with critical value $1$) and $x=1$ (with critical value $0$), the preimage of $0$ is the black vertex at $x=1$ and the preimage of $1$ consists of the white vertices at $x \in \{-1,3 \pm \sqrt{8}\}$.

Summing up, one of the simplest observable proofs of Bell's theorem is found to rely on the geometry of a {\it square} and {\it four} distinct {\it dessins} associated with it. Although we still do not know 
how these {\it dessins} are related to each other, it is quite intriguing to see that all critical points live in the extension field $\mathbb{Q}(\sqrt{2})\in \bar{\mathbb{Q}}$ of the rational field $\mathbb{Q}$. Hence, a better understanding of the properties of the group of automorphisms of this field (which is itself a subgroup of the absolute Galois group $\Gamma$) may lead to fresh insights into the nature of this important theorem of quantum physics. 

\section{A wealth of other notable point-line geometries relevant to contextuality}
\label{catalog}

{\it It is also appealing to see the failure of the EPR reality criterion emerge quite directly from the one crucial difference between the elements of reality (which, being ordinary numbers, necessarily commute) and the precisely corresponding quantum mechanical observables (which sometimes anti-commute) \cite[(a)]{Mermin1993}}.

\vspace*{.2cm}




\begin{figure}
\centering 
\includegraphics[width=5cm]{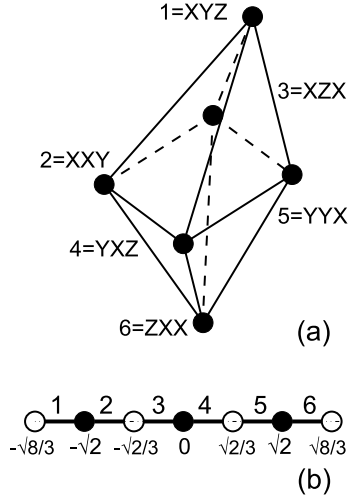}
\caption{The octahedron with vertices labeled by three-qubit observables (a) and an associated {\it dessin} (b).}
\label{fig2}
\end{figure}

\subsection{Two geometries of index six: the octahedron and the bipartite graph $K(3,3)$}

As the geometries of index five are only trivial simplices (see Table 1), we have to move to index six in order to encounter non-trivial guys, namely the octahedron and the bipartite graph of type $K(3,3)$.

\begin{figure}
\centering 
\includegraphics[width=5cm]{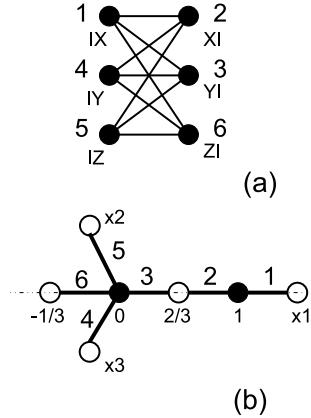}
\caption{The bipartite graph $K(3,3)$ with one of its two-qubit labelings (a) and its generating {\it dessin} (b).}
\label{fig3}
\end{figure}

\begin{figure}
\centering 
\includegraphics[width=5cm]{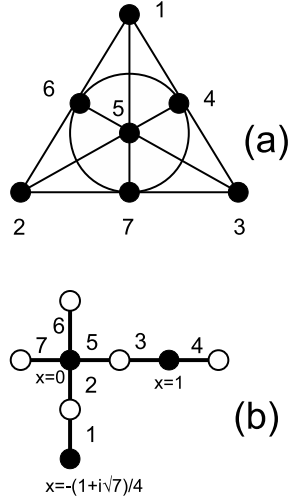}
\caption{The Fano plane portrayed in its most frequent rendering (a) and one of its ten stabilizing {\it dessins} (b).}
\label{fig4}
\end{figure}

The octahedron can be labeled by three-qubit observables, and one such labeling is given in Fig. \ref{fig2}a. The figure also illustrates one of the associated {\it dessins} (b), whose Belyi function of is $f(x)=\frac{27}{32}x^2(2-x^2)^2$. The function has critical points at $x=0$ and $x=\pm\sqrt{2}$ (these being also the preimage of $0$), and the white vertices of the {\it dessin} correspond to $x=\pm \sqrt{2/3}$ and  $x=\pm \sqrt{8/3}$ (the preimage of $1$).

A remarkable property of the graph $K(3,3)$ is that it lives in the generalized quadrangle GQ(2,\,2), disguised there as a generalized quadrangle of type GQ(1,\,2) \cite{psm}. And since GQ(2,\,2) was found to mimic the commutation relations between elements of the two-qubit Pauli group \cite{Planat2007}, $K(3,3)$ thus naturally lends itself, like the above-discussed square, to a labeling in terms of two-qubit observables --- as, for example, depicted in Fig. \ref{fig3}a. One of the associated {\it dessins} (Fig. \ref{fig3}b) possesses the Belyi function of the form $f(x)=ax^4(x-1)^2$, where $a=\frac{3^6}{2^4}=\frac{729}{16}$. Since $f'(x)=ax^3(x-1)(3x-2)$ the critical points are at $x=2/3$, $x=0$ and $x=1$. The black vertices of the {\it dessin} correspond to $x=0$ and $x=1$; out of its five white vertices three answer to real-valued  variable, namely $x=-\frac{1}{3}$, $x=\frac{2}{3}$ and $x \approx 1.118$ (the latter being denoted as $x_1$ in Fig. \ref{fig3}b), and the remaining two --- denoted as $x_2$ and $x_3$ in the figure in question --- have imaginary, complex-conjugate one: $x\approx 0.36 \exp{(\pm i \phi)}$, with $\phi \approx 99.4^{\circ}$.

\subsection{The Fano plane (everywhere)}

The only non-trivial geometry of index seven is the projective plane of order two, the Fano plane (Fig.\,\ref{fig4}a). This plane plays a very prominent role in finite-dimensional quantum mechanics, being, for example, intricately related --- through the properties of the split Cayley hexagon of order two \cite{psm} --- to the structure of the three-qubit Pauli group  \cite{Levay2008}. 
A quick computer search for all permutation subgroups of $C_2^+$ isomorphic to the group $PSL(2,7)$, the automorphism group of the Fano plane, shows that this plane can be recovered from $10$ distinct {\it dessins}. One choice is depicted schematically in Fig.\,\ref{fig4}b; it corresponds to passport $8$ (the fourth {\it dessin}) in the catalog of B\'etr\'ema $\&$ Zvonkin \cite{ZvonkinCatalog}. The corresponding permutation group is $P=\left\langle (2,7,6,5)(3,4),(1,2)(3,5)\right\rangle$ and the Belyi function reads $f(x)=\sqrt{8}x^4(x-1)^2(x-a)$, with $a=-\frac{1}{4}(1+i\sqrt{7})$;  its critical points are located at $x=0$ and $x=1$ (yielding critical value $0$) and at $x=a$ (yielding $1$).

\subsection{The $16$-cell, stellated octahedron and the completed cube}

\begin{figure}
\centering 
\includegraphics[width=6cm]{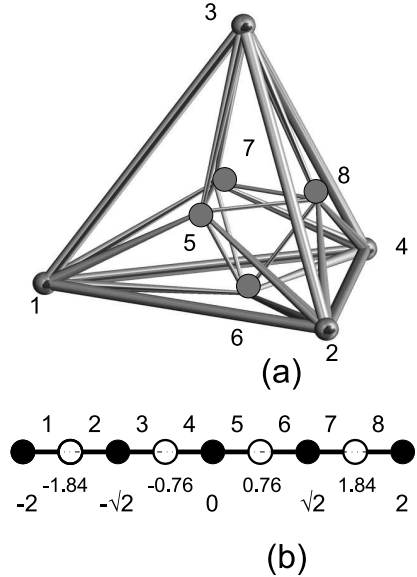}
\caption{The $16$-cell (a) and an associated {\it dessin} (b).}
\label{fig5}
\end{figure}

\begin{figure}
\centering 
\includegraphics[width=6cm]{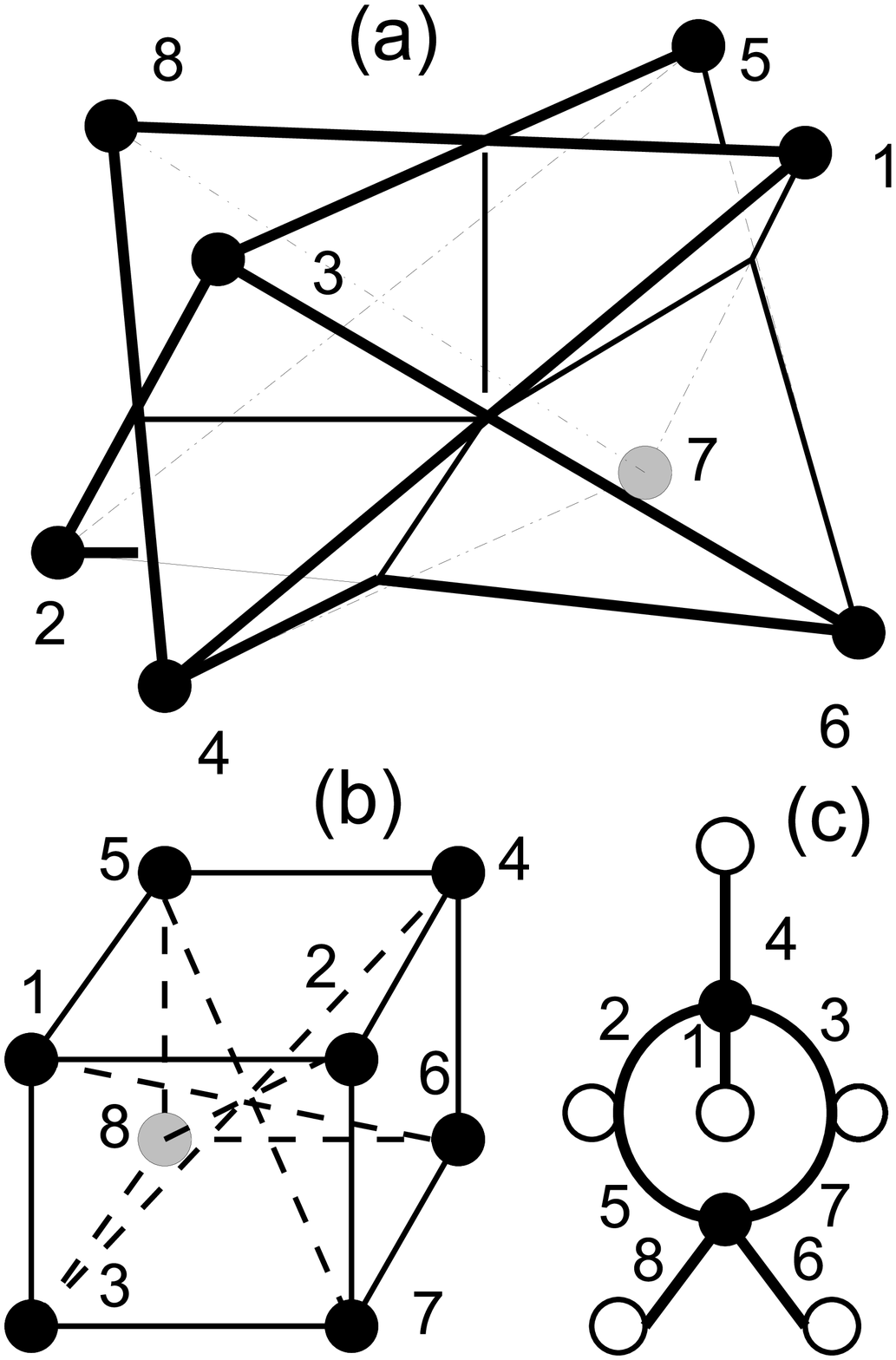}
\caption{The stellated octahedron (a), the completed cube (b) and their common stabilizing {\it dessin} (c).}
\label{fig6}
\end{figure}

When moving to index eight, we encounter an appealing 16-cell on the one hand, and the remarkably ``twinned'' stellated octahedron and completed cube on the other hand.

The $16$-cell (Fig.\,\ref{fig5}a) arises from a ``straight-line'' {\it dessin} with the signature $(5,4,1,0)$ and the permutation group isomorphic to $D_8$. Its Belyi function is the fourth order of the map $x \rightarrow x^2-2$, that is $f(x)=8x(x^2-2)(x^4-4x^2+2)$, with critical points located at $x-0$, $x=\pm \sqrt{2}$ and $x=\pm \sqrt{2 \pm \sqrt{2}}$.

The {\it dessin} with signature $s=(2,6,2,0)$ , illustrated in Fig.\,\ref{fig6}c, has the permutation group $P=\left\langle (1,2,4,3)(5,7,6,8),(2,5)(3,7)\right\rangle$, which is isomorphic to $\mathbb{Z}_2^3 \rtimes \mathbb{Z}_2$ and endowed with the cycle structure of the form $[4^2,2^2 1^4,4^2]$. The stabilizer of a pair of its edges is either the group $\mathbb{Z}_2$, leading to the geometry of a stellated octahedron (Fig.\,\ref{fig6}a), or the trivial single element group $\mathbb{Z}_1$, in which case we get the geometry of a (triangle free) {\it completed cube}, i.\,e. the ordinary cube where pairs of opposite points are joined (Fig.\,\ref{fig6}b). The latter configuration also appears in a recent paper \cite[pp. 33--34 as well as Conjecture 6.1]{Grunbaum} as an $8$-face Kepler-Poinsot quadrangulation of the torus. Note that, in addition the $6$ faces shared with the ordinary cube, the completed cube contains also $8$ non-planar faces of which four are self-intersecting.  The completed cube can also be viewed as the bipartite graph $K(4,4)$. The Belyi function of the {\it dessin} has the form
$$f(x)=K\frac{(x-1)^4(x-a)^4}{x^3},~~a=\frac{8\sqrt{10}-37}{27},~~K \approx-0.4082,$$
from where we find the positions of  ``critical'' white vertices to be $x \approx 0.0566 \pm 0.506 i$, with the other four white vertices being located at $x=-1.069$, $x=-0.162$ and $x=1.634 \pm 0.6109i$.

\subsection{Geometries of index nine: grid (Mermin's square), Pappus and Hesse}

In the realm of index nine we meet, in addition to our old friend, a $3 \times 3$-grid (alias generalized quadrangle GQ(2,\,1)), also other distinguished finite geometries like the Pappus $9_3$-configurations and the Hesse $9_{4}12_{3}$-configuration ({\it aka} the affine plane of order three, AG(2,\,3)). 

\begin{figure}[ht]
\centering 
\includegraphics[width=6cm]{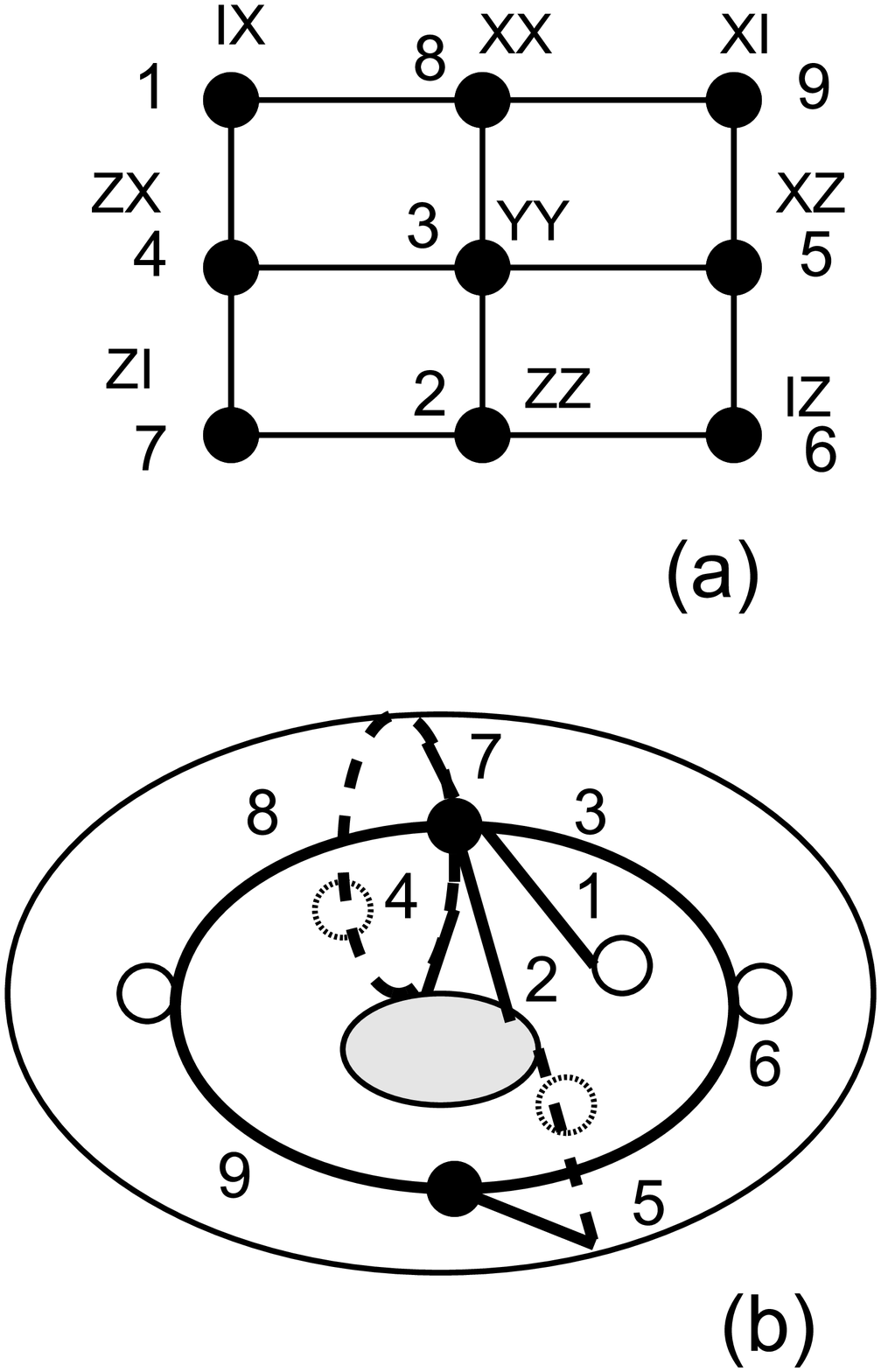}
\caption{A $3\times 3$ grid with points labeled by two-qubit observables ({\it aka} a Mermin magic square) (a) and a stabilizing {\it dessin} drawn on a torus (b).}
\label{fig7}
\end{figure}

\begin{figure}
\centering 
\includegraphics[width=6cm]{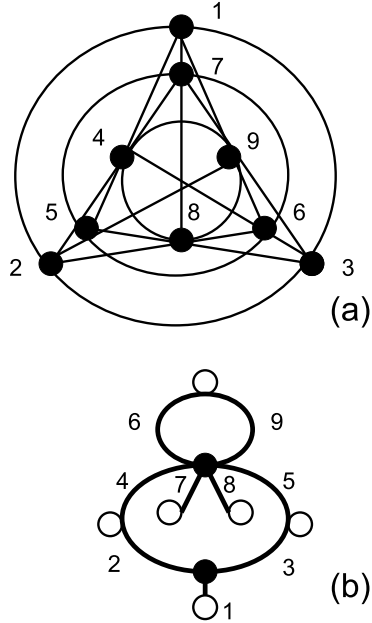}
\caption{The Hesse configuration (a) and an associated genus-zero {\it dessin} (b).}
\label{figHesse}
\end{figure}

\begin{figure}
\centering 
\includegraphics[width=6cm]{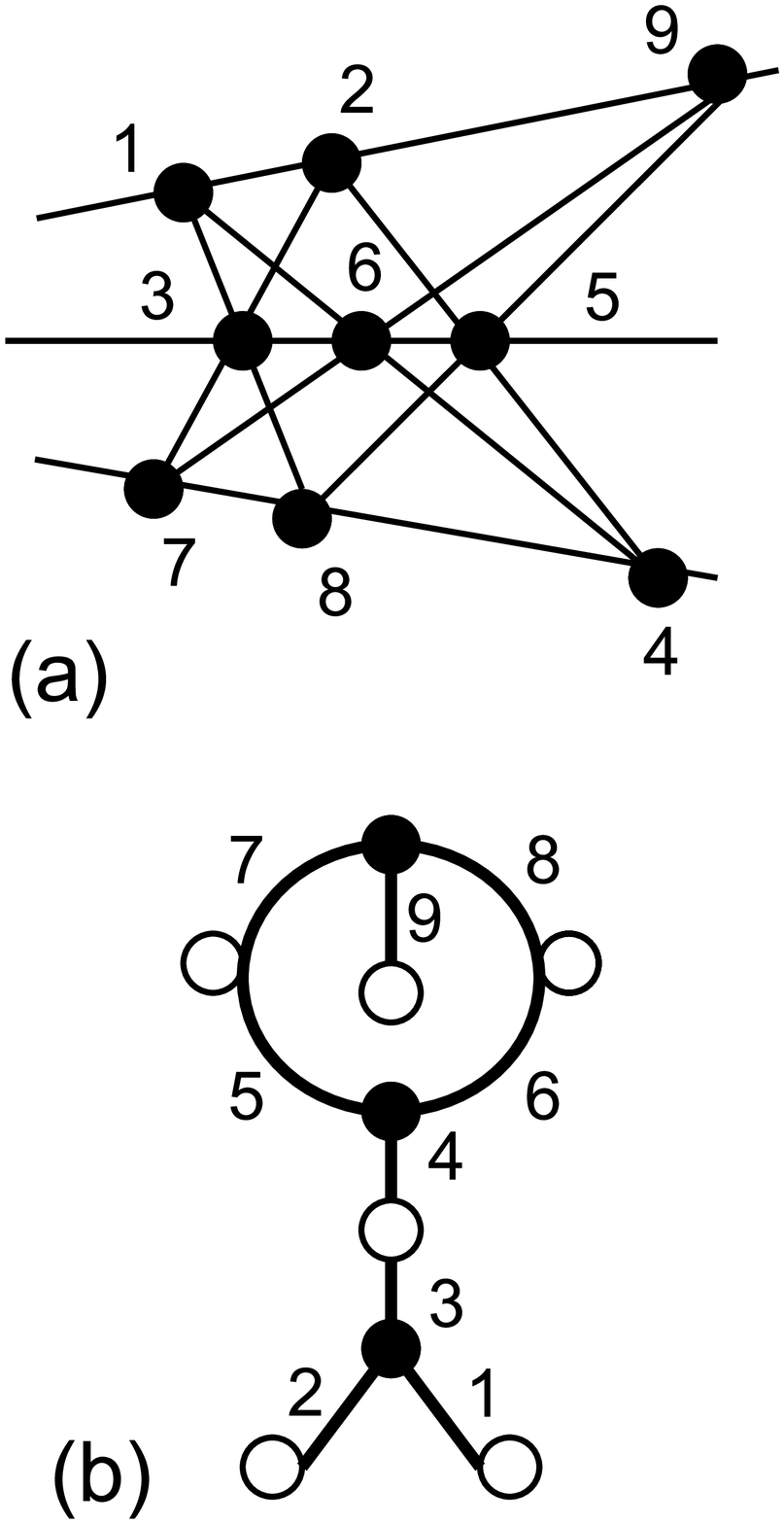}
\caption{The Pappus configuration (a) and a stabilizing {\it dessin} (b).}
\label{fig8}
\end{figure}

As already mentioned, the grid lives (as a geometric hyperplane) in GQ(2,\,2) and underlies a Mermin ``magic'' square array of observables furnishing a simple two-qubit proof of the Kochen-Specker theorem \cite{PlanatFQXi,Planat2013bis}. A Mermin square built around Bell's square of Fig. \ref{fig1}a is shown in Fig. \ref{fig7}a. It needs a genus one {\it dessin}, with signature $(2,5,2,1)$, to be recovered, as shown in Fig. \ref{fig7}b. The corresponding permutation group is 
$P=\left\langle (1,2,4,8,7,3)(5,9,6),(2,5)(3,6)(4,7)(8,9)\right\rangle \cong \mathbb{Z}_3^2 \rtimes \mathbb{Z}_2^2$, having the cycle structure  $[6^13^1,2^41^1,6^13^1]$.
This {\it dessin} lies on a Riemann surface that is a torus (not a sphere $\hat{\mathbb{C}}$), being thus represented by an elliptic curve. The topic is far more advanced and we shall not pursue it in this paper (see, e.\,g., \cite{Girondo} for details).
The stabilizer of a pair of edges of the {\it dessin} is either the group $\mathbb{Z}_2$, yielding Mermin's square $M_1$ shown in Fig \ref{fig7}a, or the group $\mathbb{Z}_1$, giving rise to a different square $M_2$ from the maximum sets of mutually non-collinear pairs of points of $M_1$. The union of $M_1$ and $M_2$ is nothing but the Hesse configuration.

The Hesse configuration (Fig.\,\ref{figHesse}a), of its own, can be obtained from a genus-zero {\it dessin}  shown in Fig.\,\ref{figHesse}b (also reproduced in Fig.\,3b of \cite{PlanatFQXi}). This configuration was already noticed to be of importance in the derivation of a Kochen-Specker inequality in \cite{Bengtsson}.

The Pappus configuration, illustrated in Fig.\,\ref{fig8}a, comprises three
copies of the already discussed $K(3,3)$-configuration  (Fig.\,\ref{fig3}a); the three copies are represented by the point-sets $\{ 1,3,5,6,7,8\}$, $\{ 2,3,4,5,8,9\}$ and  $\{1,2,4,6,7,9\}$, which pairwise overlap in distinct triples of points. A {\it dessin d'enfants} for the Pappus configuration is exhibited in Fig.\,\ref{fig8}b.
It is important to point out here a well-known fact that the Pappus configuration is obtained from the Hesse one by removing three mutually skew lines from it (for example, the three lines that are represented in Fig.\,\ref{figHesse}a by three concentric circles).

%

\subsection{Realm of index ten: Mermin's pentagram, Petersen and Desargues}

Apart from the two plexes (see Table 1), the only connected configurations generated by $10$-edge {\it dessins} are Mermin's pentagram, the Petersen graph, the Desargues configuration and the bipartite graph $K(5,5)$.

\begin{figure}[ht]
\centering 
\includegraphics[width=7cm]{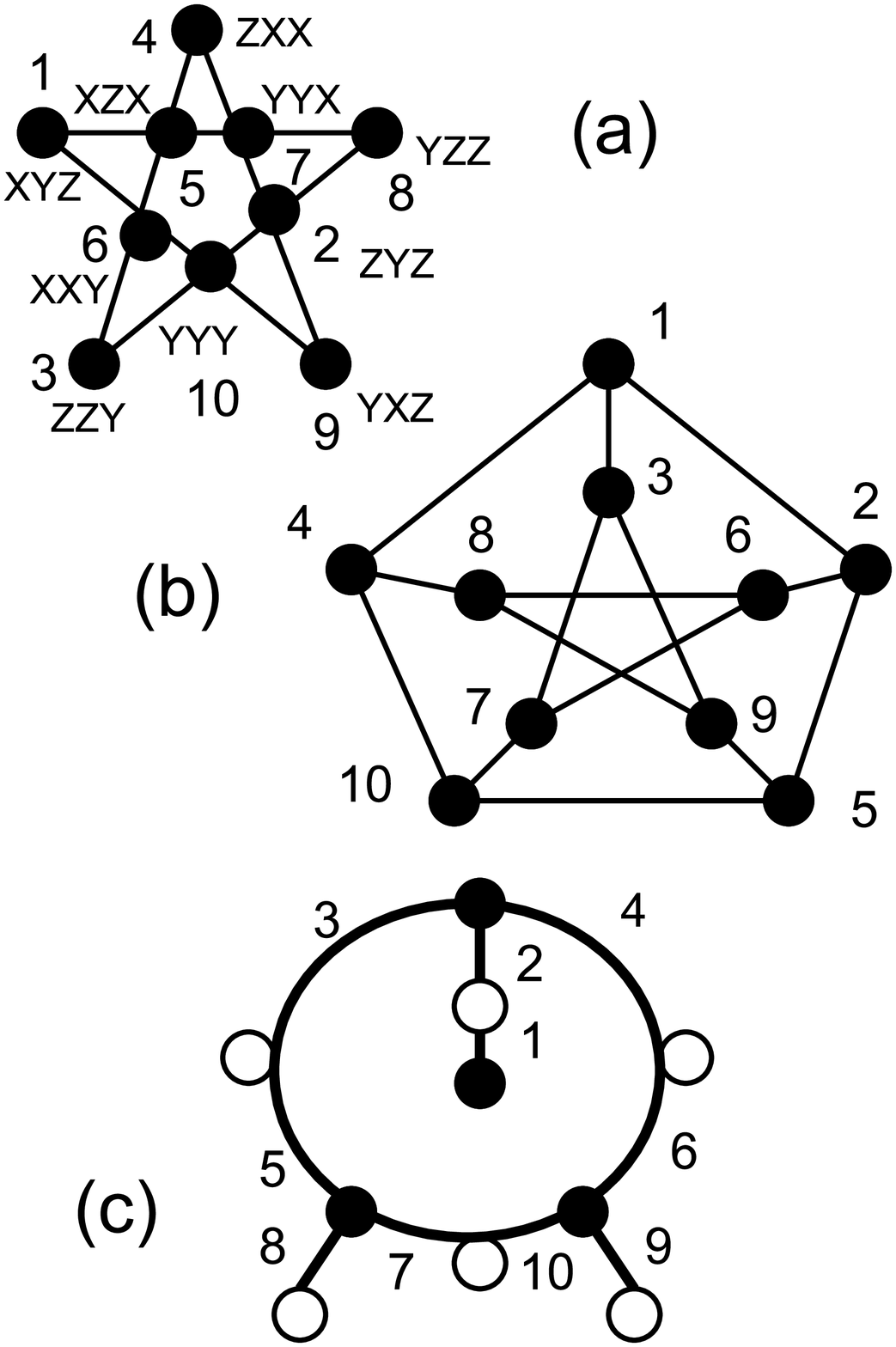}
\caption{The Mermin pentagram (a), the Petersen graph (b) and their generating {\it dessin} (c).}
\label{fig9}
\end{figure}

The {\it dessin} sketched in  Fig.\,\ref{fig9}c, having $s=(4,6,2,0)$ and the alternating group $A_5$ with cycle structure $[3^21^1,2^41^2,5^2]$, induces either the geometry of Mermin's pentagram (Fig.\,\ref{fig9}a) or that of the Petersen graph (Fig.\,\ref{fig9}b) according as the group stabilizing pairs of its edges is isomorphic to $\mathbb{Z}_1$ or $\mathbb{Z}_2$, respectively. A particular three-qubit realization leading to a proof of the Kochen-Specker theorem is explicitly shown (see, e.\,g., \cite{Planat2013, Planat2013bis} for more details on importance of these geometries in quantum theory).

\begin{figure}
\centering 
\includegraphics[width=7cm]{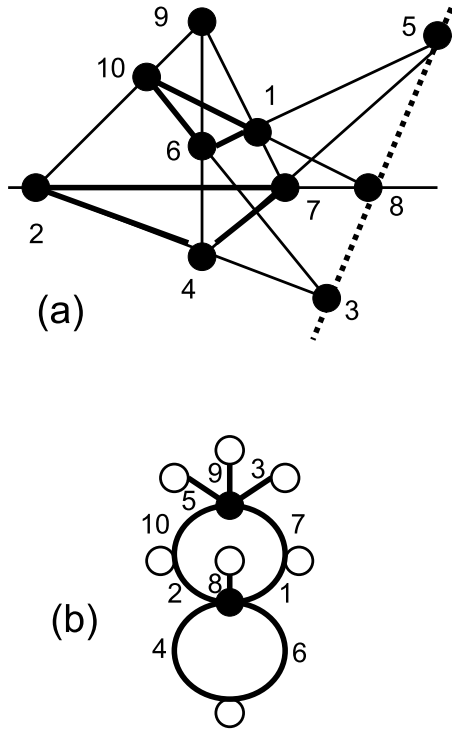}
\caption{The Desargues configuration (a) and its generating dessin (b).}
\label{fig10}
\end{figure}

The {\it dessin} depicted in Fig.\,\ref{fig10}b also gives rise to a couple of geometries,  one being  again the Petersen graph (with the stabilizer group of a pair of edges isomorphic to $\mathbb{Z}_2^2$) and the other being (Fig.\,\ref{fig10}a) the famous Desargues $10_3$ configuration  (with the group $\mathbb{Z}_2$). The labeling is compatible with that in Fig.\,\ref{fig9}, which means that the Desargues configuration represents another way of encoding a three-qubit proof of contextuality; in particular, a line of Mermin's pentagram corresponds to a complete graph $K_4$ within the Desargues configuration as well as to a maximum set of mutually disjoint vertices in the Petersen graph.

\begin{figure}[ht]
\centering 
\includegraphics[width=8cm]{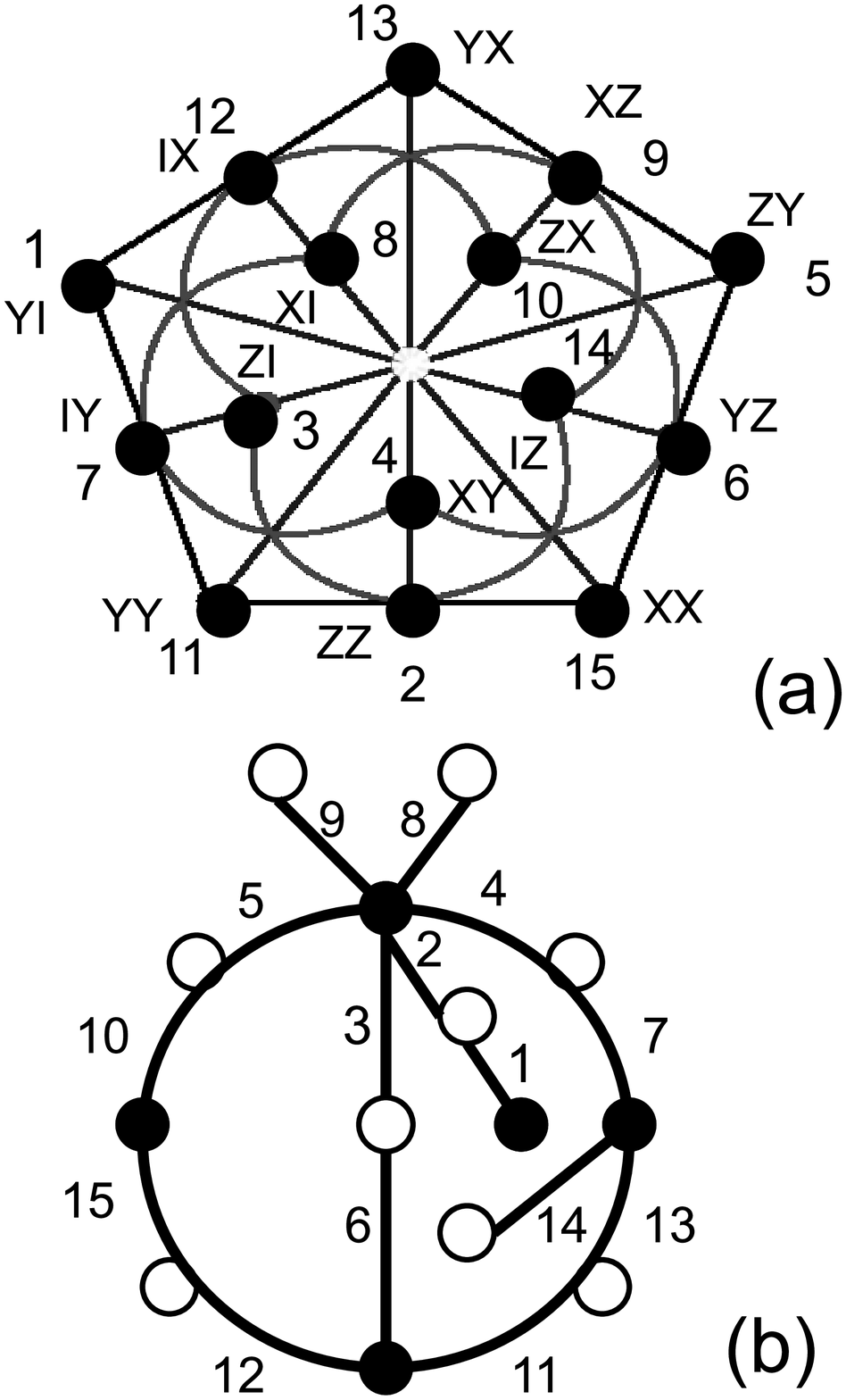}
\caption{The Cremona-Richmond $15_3$-configuration (a) with its points labeled by the elements of the two-qubit Pauli group and a stabilizing {\it dessin} (b).}
\label{fig11}
\end{figure}

\subsection{The Cremona-Richmond $15_3$-configuration, alias GQ(2,\,2), or W(3,\,2)}

We now come to a perhaps most exciting, and encouraging as well, finding that
there exists a {\it dessin} generating the configuration of a central importance for any quantum physical reasoning involving two-qubit observables, namely the configuration (illustrated in Fig.\,\ref{fig11}a) known as $15_3$ Cremona-Richmond configuration, or the generalized quadrangle of order two, GQ(2,\,2), or the symplectic polar space of rank two and order two, W(3,\,2). That this configuration is indeed one of corner-stones of finite dimensional quantum mechanics is also illustrated by the fact that many of the already discussed geometries, in particular the $K(3,3)$ graph, the $3 \times 3$ grid, the Pappus and Desargues configurations and the Petersen graph, are intricately tied to its structure, as explained in detail in \cite{Planat2007},\cite{spph2007}--\cite{hos2009}. The associated {\it dessin} (Fig.\,\ref{fig11}b) is of signature $(5,9,3,0)$ and its permutation group has the cycle structure $[6^13^22^11^1,2^61^3,6^23^1]$. Unfortunately, the complexity of this {\it dessin} is already so high that with our current computer power we have not been able to find the corresponding Belyi function. Finding this function thus remains one important challenge of our {\it dessin d'enfants} programme.



\subsection{The generalized quadrangle GQ(2,\,4), the Schl\"{a}fli graph and the Clebsch graph}
\begin{figure}
\centering 
\includegraphics[width=6cm]{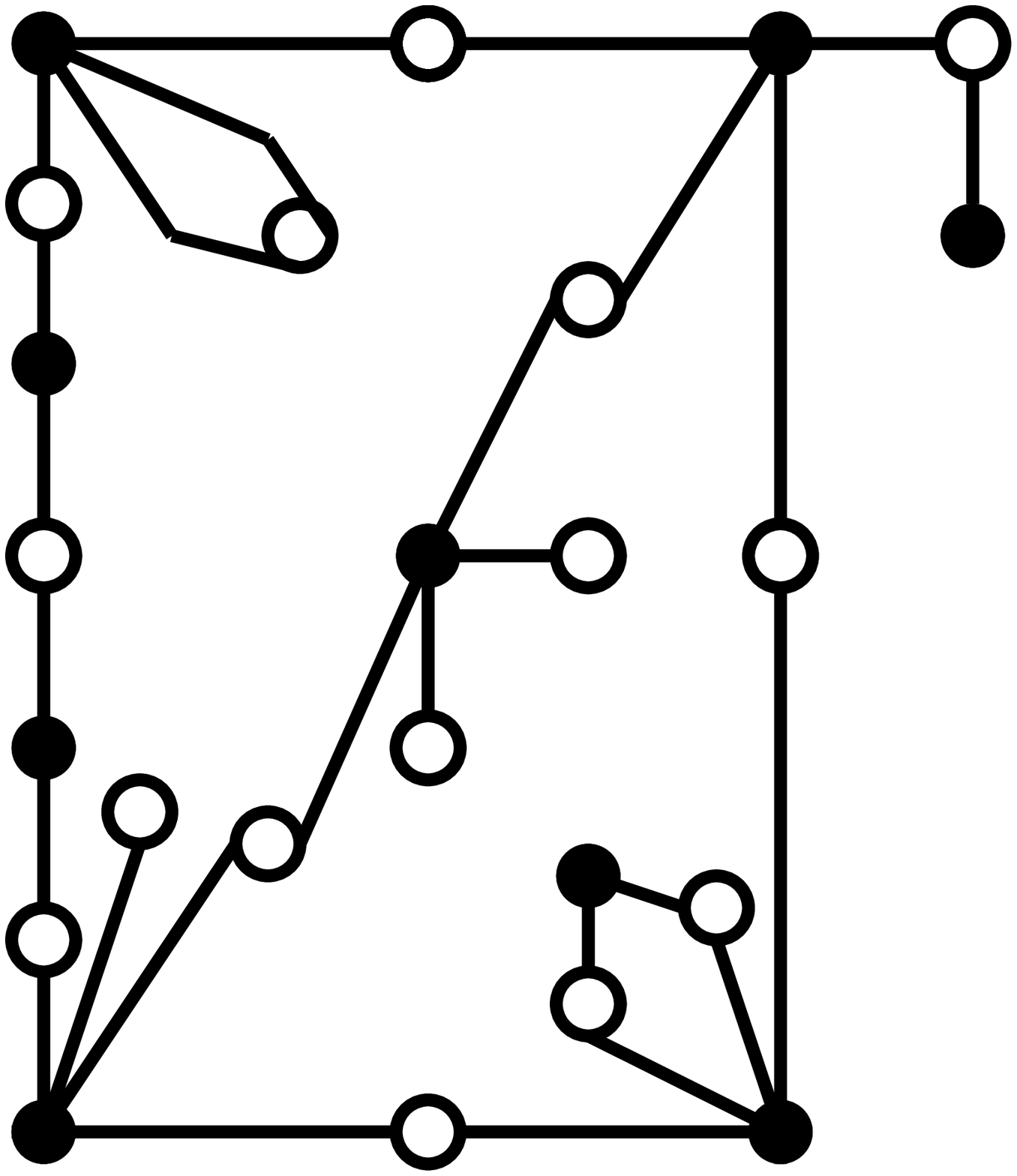}
\caption{The {\it dessin} for both the collinearity graph of the generalized quadrangle of type GQ(2,\,4) and its complement, the Schl\"{a}fli graph. }
\label{fig12}
\end{figure}

As already mentioned, the generalized quadrangle of type GQ(2,\,4) is a prominent finite geometry in the context of the so-called black-hole--qubit correspondence \cite{Levay2009}, as it completely underlies the $E_6$-symmetric entropy formula describing black holes and black strings in $D = 5$. We were thus very pleased to find a {\it dessin} that leads to the collinearity graphs of this geometry, and to its complement -- the famous Schl\"{a}fli graph -- as well (Fig.\,\ref{fig12}).  Moreover, GQ(2\,,4) is also notable by the fact it contains (altogether 27) copies of the Clebsch graph, each such copy being the complement of a geometric hyperplane of particular type. And since the Clebsch graph is also a {\it dessin}-generated one (see Table\,2), we will not be surprised if our {\it dessin}'s formalism is also found of relevance for getting conceptual insights into the still-mysterious formal link between stringy black hole entropy formulas and properties of multi-qubits (for a recent review, see \cite{bdl}).   

\section{Conclusion}

We have demonstrated, substantially boosting the spirit of \cite{PlanatFQXi},
that Grothendieck's {\it dessins d'enfants} (``child's drawings'') --- that is graphs where at each vertex is given a cyclic ordering of the edges meeting it and each vertex is also assigned one of two colors, conventionally black and white, with the two ends of every edge being colored differently --- and their associated permutation groups/Belyi functions give rise to a wealth of finite geometries relevant for quantum physics.
We have made a complete catalog of these geometries for {\it dessins} featuring up to 12 edges, highlighted distinguished geometries for some higher-index {\it dessins}, and briefly elaborated on quantum physical meaning for each non-trivial geometry encountered. We are astonished to see that a majority of dessin-generated geometries have already been found to have a firm footing in finite-dimensional quantum mechanical setting, like the K(3,\,3) and Petersen graphs, the Fano plane, the $3 \times 3$ grid (Mermin's square), the Desargues configuration, Mermin's pentagram and the generalized quadrangles GQ(2,\,2) and GQ(2,\,4). We have also found a wealth of geometries, among them the Hesse $9_4 12_3$-configuration, the Reye  $12_4 16_3$-configuration \cite{Aravind}, the $3 \times 3 \times 3$-grid, the Kneser graph $KG_{(7,2)}$ and many others, that still await their time to enter the game. Our findings may well be pointing out that properties of {\it dessins}, as well as the Galois group $\mathcal{G}=\mbox{Gal}(\bar{\mathbb{Q}}/\mathbb{Q})$ acting on them, may be vital for getting deeper insights into foundational aspects of quantum mechanics. 
To this end in view, we aim to expand in a systematic way our catalog of finite geometries generated by higher-index dessins in order to reveal finer traits of the quantum pattern outlined above.



\section*{Acknowledgments}

This work started while two of the authors (M.\,P. and M.\,S.) were fellows of the ``Research in Pairs'' Program of the Mathematisches Forschungsinstitut Oberwolfach (Oberwolfach, Germany), in the period from 24 February to 16 March, 2013. It was also supported in part by the VEGA Grant Agency, grant No. 2/0003/13.


\end{document}